\documentclass[useAMS,usenatbib]{mn2e}
\pdfoutput=1
\usepackage{appendix}
\usepackage{amssymb}
\usepackage[pdftex]{graphicx}
\usepackage[pdftex,usenames,dvipsnames]{xcolor}
\usepackage{natbib}
\usepackage{aas_macros}
\usepackage{url}
\usepackage[none]{hyphenat}
\usepackage{times}
\usepackage[linktocpage=true,backref=true,pagebackref=false,bookmarks=true,bookmarksnumbered=False,colorlinks=true,linkcolor=NavyBlue,citecolor=NavyBlue,urlcolor=NavyBlue]{hyperref}
\usepackage[all]{hypcap}
\usepackage[color = white,bordercolor = ForestGreen]{todonotes}
\pdfpageattr {/Group << /S /Transparency /I true /CS /DeviceRGB>>}  

\title[The resolved structure of  SNR 4449--1]{The resolved structure of the extragalactic supernova remnant SNR\,4449-1}
\author[M. Mezcua et al.]
  {M.~Mezcua,$^1$\thanks{Email: mmezcua@iac.es\newline Present address: Instituto de Astrof\'isica de Canarias, V\'ia L\'actea S/N, La Laguna 38200, Tenerife, Spain}
  A.P.~Lobanov,$^1$\thanks{Visiting Scientist, University of Hamburg / Deutsches Elektronen Synchrotron Forschungszentrum.}
  I.~Mart\'i-Vidal,$^{1,2}$ \\
  $^1$Max Planck Institute for Radio Astronomy,
              Auf dem H\"ugel 69, D-53121 Bonn, Germany \\
  $^2$Onsala Space Observatory, Chalmers University of Technology, Sweden }

\date{Accepted 2013 September 13}

\pagerange{\pageref{firstpage}--\pageref{lastpage}} \pubyear{2013}

\def\LaTeX{L\kern-.36em\raise.3ex\hbox{a}\kern-.15em
    T\kern-.1667em\lower.7ex\hbox{E}\kern-.125emX}

\begin{document}

\label{firstpage}

\maketitle

\begin{abstract}
We present very long baseline interferometry (VLBI) observations of the
milliarcsecond-scale radio structure of the supernova remnant SNR 4449$-$1 in the galaxy NGC
4449. This young and superluminous remnant was observed at 1.6 GHz
($\lambda = 18$\,cm) with the European VLBI Network. The
observations confirm earlier identifications of this object with a
supernova remnant (SNR) while revealing a somewhat different morphology compared with the
structure reported by Bietenholz et al. from VLBI observations
at 1.4~GHz. This difference is discussed here in the context of
structural sensitivity of both observations. The 1.6~GHz image yields
accurate estimates of the size (0.0422 arcsec $\times$
0.0285 arcsec and 0.8 $\times$ 0.5 pc) and age ($\sim$55
yr) of SNR~4449$-$1. With a total flux of 6.1 $\pm$ 0.6 mJy
measured in the VLBI image, the historical lightcurve of the source
can be well represented by a power-law decay with a power index of
$-$1.19 $\pm$ 0.07.  The SNR exhibits a decline rate of the radio
emission of 2.2$\%$ $\pm$ 0.1$\%$ yr$^{-1}$ and a radio luminosity of
1.74 $\times$ 10$^{35}$ erg s$^{-1}$.
 \end{abstract}


\section{Introduction}

The young and ultraluminous supernova remnant SNR
4449$-$1 
was discovered in the late 1970s (radio,
\citealt{1978ApJ...226L...5S}; optical, \citealt{1978ApJ...226L...7B})
in the star-forming galaxy NGC\,4449.  NGC\,4449 is a barred
Magellanic-type irregular galaxy hosting a number of areas of
extensive star formation (\citealt{2008AJ....135.2222R}). The galaxy
is located at an estimated distance of 3.82 $\pm$ 0.27 Mpc
(\citealt{2008AJ....135.1900A}).

The SNR 4449$-$1 was identified as a young and very luminous
oxygen-rich supernova remnant (SNR) embedded in an HII region
(\citealt{1998AAS...193.7404B}), based on observations in several
different wavebands.  It was first discovered as a bright, unresolved
non-thermal radio source ($\sim$10 mJy at 2.7 GHz;
\citealt{1978ApJ...226L...5S}) located approximately 1arcmin north of the nucleus of NGC\,4449.  A steep spectral index
$\mathrm\alpha = -0.95 \pm 0.35$\footnote{The spectral index $\alpha$
  is defined here from $S_{\nu} \propto \nu^{\alpha}$, where
  \textit{S} is the flux density at frequency $\nu$}
(\citealt{1978ApJ...226L...5S}) in the radio band and observations of
both broad and narrow lines in the optical spectrum
(\citealt{1978ApJ...226L...7B}) indicated that the source was a SNR.

SNR 4449$-$1 is one of the few known intermediate-age SNRs, with
an age between that of Cas A ($\sim$330 yrs;
\citealt{2001AJ....122..297T}) and those of the oldest known
extragalactic radio supernovae (SNe), like SN 1923A in M83 (83 yrs;
\citealt{1998ApJ...508..664E}) and SN 1957D, also in M83 (49 yrs;
\citealt{1983ApJ...270L...7P}).  The SNR in NGC 4449 is also notable
as the most luminous and most distant member of the class of
oxygen-rich SNRs (\citealt{1983ApJ...272...84B}). The nature of such a
bright luminosity was long suspected to be due to interaction with a
surrounding HII region, but \textit{Hubble Space Telescope} (\textit{HST}) observations
(\citealt{2008ApJ...677..306M}) suggested that the remnant is instead
interacting with a very dense circumstellar material (CSM) from the SNR's
progenitor star of mass $\geq$ 20 M$\odot$. The remnant possesses the
characteristics of typical SNe of massive stars, since it lies inside
a rich cluster of high-mass stars surrounded by a presumably SN- and
wind-blown interstellar medium bubble, it is interacting with
the dense CSM and it exhibits the chemical
properties of an hydrogen-poor envelope progenitor
(\citealt{2008ApJ...677..306M}).

While SNR 4449$-$1 is currently quite luminous, the significant
decline in the remnant's X-ray and radio flux over the last three
decades implies that it was even brighter in the
past. \cite{2007AJ....133.2156L} combined newer Very Large Array (VLA)
observations of NGC 4449 with archive VLA and Westerbork Synthesis
Radio Telescope observations to present the lightcurve of the
SNR at 6 and 20 cm from 1973 to 2002.  The radio flux of SNR
4449$-$1 measured at 4.9 GHz has undergone a signficant decline over
the last three decades, dropping from 13 to 4\,mJy between 1973
and 2002 (\citealt{2007AJ....133.2156L}).

SNR 4449$-$1 was identified with an ultraluminous X-ray
source (ULX), NGC 4449-X4, based on \textit{ROSAT} High Resolution Imager (HRI)
observations (e.g., \citealt{2005ApJS..157...59L};
\citealt{2006A&A...452..739S}). Application of an absorbed,
non-equilibrium ionization model (e.g., \citealt{2002PASJ...54...53Y})
to NGC 4449-X4 yielded an X-ray temperature $T_\mathrm{x} \approx 2.2
\times 10^{7}$\,K in the 0.3--8.0 keV band, a column density of
$N_\mathrm{H}$ = 1.26 $\times$ 10$^{21}$ cm$^{-2}$ and an
absorption-corrected luminosity $L_\mathrm{x} = 2.3 \times
10^{38}$\,erg s$^{-1}$ (\citealt{2003MNRAS.342..690S}). The X-ray
temperature measurement was used to estimate an age of $\sim$270 yrs
and an ambient medium density of 120--200 cm$^{-3}$ for this SNR
(\citealt{2003MNRAS.342..690S}).

Very long baseline interferometry (VLBI) observations made in 1980
  and 1981 with the EVN\footnote{The European VLBI Network,
  www.evlbi.org} yielded an upper limit of $\leq$
0.07 arcsec (1.3 pc at 3.82 Mpc) for the angular diameter of
SNR 4449$-$1 (\citealt{1983A&A...119..301D}). This upper limit was
obtained from interferometric visibilities, while the data were
insufficient for obtaining an image. Optical observations with the
\textit{HST} in 1996 and 2005 yielded
somewhat smaller upper limits of 0.028 arcsec (0.5 pc; 
\citealt{1998AAS...193.7404B}) and 0.037 arcsec
(0.69 pc; \citealt{2008ApJ...677..306M}), respectively. The expansion velocity
of 6000 km s$^{-1}$ obtained from the [OIII] optical emission line
(\citealt{1998AAS...193.7404B}) provided an estimated age of the
remnant of $\sim$ 50 yr (\citealt{2008ApJ...677..306M}).  Recently,
\cite{2010MNRAS.409.1594B} reported the first resolved radio image of
the remnant, obtained with VLBI and showing two parallel ridge-like
structures with an angular extent of 65 mas $\times$ 40 mas. Using a
slightly larger velocity expansion of 6500 km s$^{-1}$, they obtained
an age estimate of $\sim$ 70 yrs for the SNR.

We have made a VLBI observation of SNR 4449$-$1 using the EVN at
1.6 GHz ($\lambda = 18$\,cm), as part of a larger study of
milliarsecond-scale structure of radio counterparts of ULXs
(\citealt{2011AN....332..379M}; \citealt{2013arXiv1309.4463M}a) identified from a cross-correlation of
positions of ULX objects with the FIRST\footnote{Faint Images of the Radio Sky at Twenty-cm} catalogue
(\citealt{2006A&A...452..739S}). General results of this study will be
described in a forthcoming paper, and here we focus on discussing
the properties of milliarcsecond-scale emission in SNR 4449$-$1.

In Section~2, we describe the EVN observations and data reduction.
The resulting image of the millarcsecond-scale structure of radio
emission from SNR 4449$-$1 is presented in Section~3. A discussion
given in Section~4 brings these results in the context of long-term
evolution of the SNR.  We adopt a distance to the SNR of 3.82 Mpc
(\citealt{2008AJ....135.1900A}), which corresponds to a linear scale
of 18.4 pc arcsec$^{-1}$.

\section{Observations and data reduction}

SNR 4449$-$1 in NGC\.4449 was observed during a 12 h observing
run, during which two other ULXs were also observed, on
2009 June 1 using the EVN at 1.6 GHz (wavelength of
18\,cm). Nine EVN antennas participated in the observations, and their
basic technical parameters are listed in Table~\ref{tb:evnobs}.

The observations were made in the phase-referencing mode, with the
compact radio source J1221+4411, located $\sim$1$^{\circ}$ away from the
target, used as a phase calibrator. Observing scans on J1221+4411 and
SNR 4449--1 were interleaved, with a calibrator--target cycle time of 8
min (2 min on the phase calibrator and 6 min on the source). This
resulted in a total integration time of 2 h spent on SNR 4449--1.
A bright and relatively compact radio source, 4C\,+39.25, was observed as a
fringe finder and bandpass calibrator.

The data were recorded in dual-circular polarization, at a rate of
1024 megabit per second (Mbps), resulting in a total bandwidth of 256
MHz (comprising eight intermediate frequency  bands, each covering a 16
MHz band split into 32 spectral channels). After the observations, the
data were correlated at the EVN Correlator Facility at
JIVE\footnote{Joint Institute for VLBI in Europe, Dwingeloo, the
Netherlands.} using a correlator averaging time of 2s. This
ensured that the target source could be effectively detected within
the 3$\sigma$ error box of the VLA FIRST position measurement.

The correlated data were fringe-fitted using the
AIPS\footnote{Astronomical Image Processing Software of National Radio Astronomy
Observatory (NRAO).} and
phase solutions obtained on the phase-reference calibrator J1221+4411
were applied to the target source.  No ionospheric corrections were
applied. The target was subsequently imaged using the AIPS task IMAGR. The
resulting image is shown in Fig.~\ref{fig1} and discussed below.

\begin{table}
\caption{Technical characteristics of the EVN observations}
\label{tb:evnobs}
\begin{center}
\begin{tabular}{lcrc}\hline\hline
\multicolumn{4}{c}{Participating telescopes} \\\hline
Antenna           & $D$ & SEFD    & $\sigma_\mathrm{n}$ \\
                  & (m) & (Jy)    & (mJy)     \\\hline
Effelsberg (DE)   & 100 & 19      &  ...      \\
Jodrell Bank (UK) & ~76  & 44      &  0.24   \\
Medicina (IT)     & ~32  & 600     &  0.87     \\
Noto (IT)         & ~32  & 780     &  1.00     \\
Onsala (SE)       & ~25  & 320     &  0.64     \\
Shanghai (CN)     & ~25  & 670     &  0.92     \\
Torun (PL)        & ~32  & 230     &  0.54     \\
Urumqi (CN)       & ~25  & 270     &  0.58     \\
Westerbork (NL)   & ~~83$^\mathrm{a}$  & 30      &  0.20     \\
\hline
\end{tabular}
\end{center}
Notes.~$D$ -- antenna diameter; SEFD -- system equivalent flux
density (an integral measure of antenna sensitivity);
$\sigma_\mathrm{n}$ -- rms noise for 1 min integration on the
baseline between the given antenna and Effelsberg; $^\mathrm{a}$ -- equivalent
antenna diameter for a phased array of 11$\times$25 m antennas
used for the observations.
\end{table}

\section{Results}

\begin{figure}
\centering
\includegraphics[width=\columnwidth]{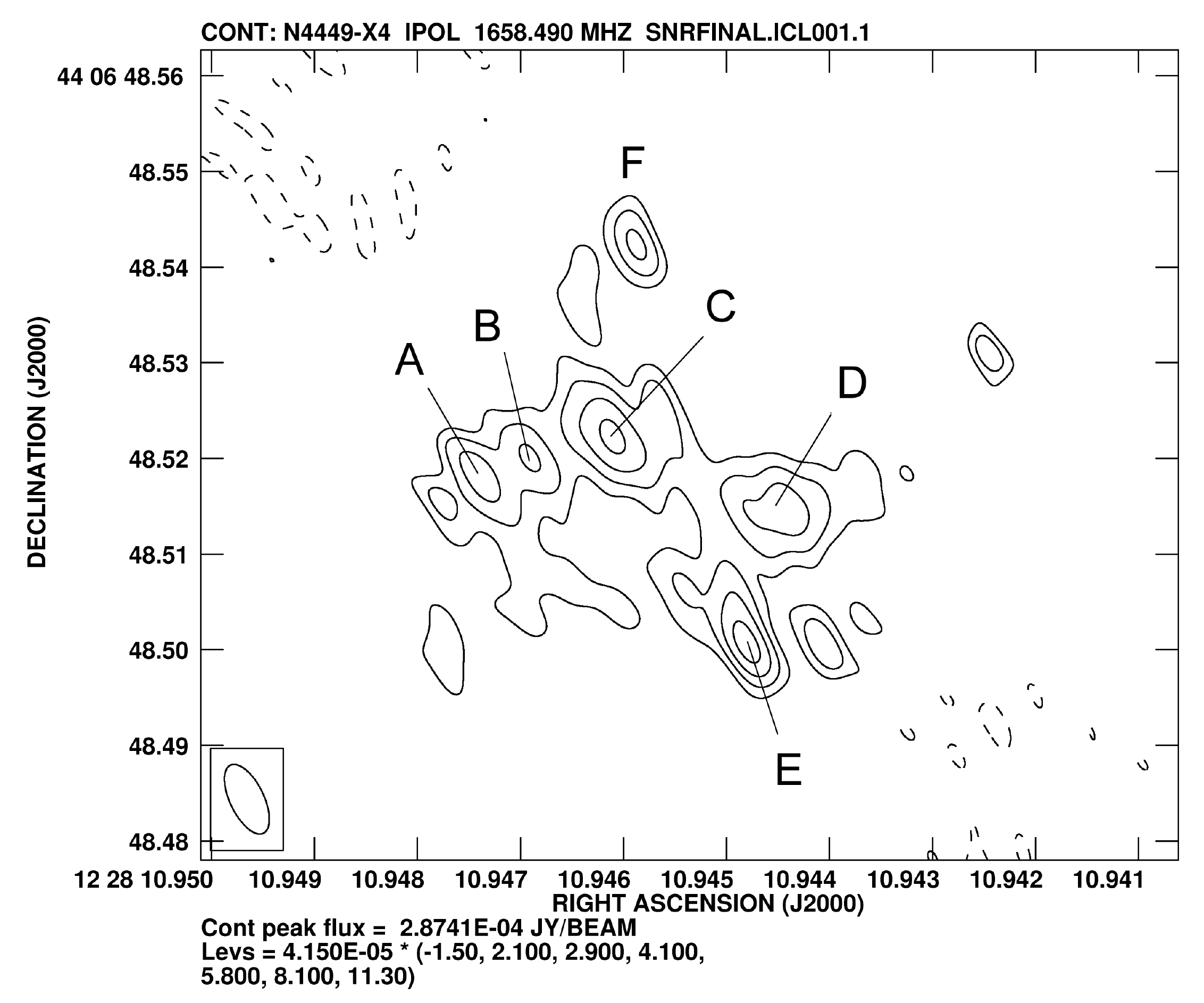}
\protect\caption{Resolved radio structure of the SNR in NGC 4449 using the EVN at 1.6 GHz. The rms
noise off-source is $42\mu$\,Jy\,beam$^{-1}$. Contours start at $-1.5$ times the rms and 
increase with factors of $\sqrt{2}$. The six components detected are labelled A, B, C, D, E and F. The brightness peak of the map corresponds to component E and is $0.287$\,mJy\,beam$^{-1}$. 
The dimensions (full width at half-maximum) of the restoring beam are $7.9$\,mas$\times3.5$\,mas, with 
the major axis of the beam oriented along a position angle of 25$^{\circ}$.8. North is up and 
east is to the left.}
\label{fig1}
\end{figure}

A phase-referenced image of SNR 4449--1 shown in Fig.~\ref{fig1}
is obtained by applying CLEAN deconvolution to the naturally weighted
data. The cleaning is done without self-calibration.  The resulting
restoring beam (CLEAN beam) is 7.9 mas $\times$ 3.5 mas, oriented at a
position angle of 25$^{\circ}$.8. The image has an off-source rms noise of
$42\,\mu$Jy beam$^{-1}$ and a peak flux density of $287\,\mu$Jy
beam$^{-1}$, thus corresponding to a detection of signal-to-noise
ratio (S/N)$\approx 7$.

\begin{figure}
\centering
\includegraphics[width=\columnwidth]{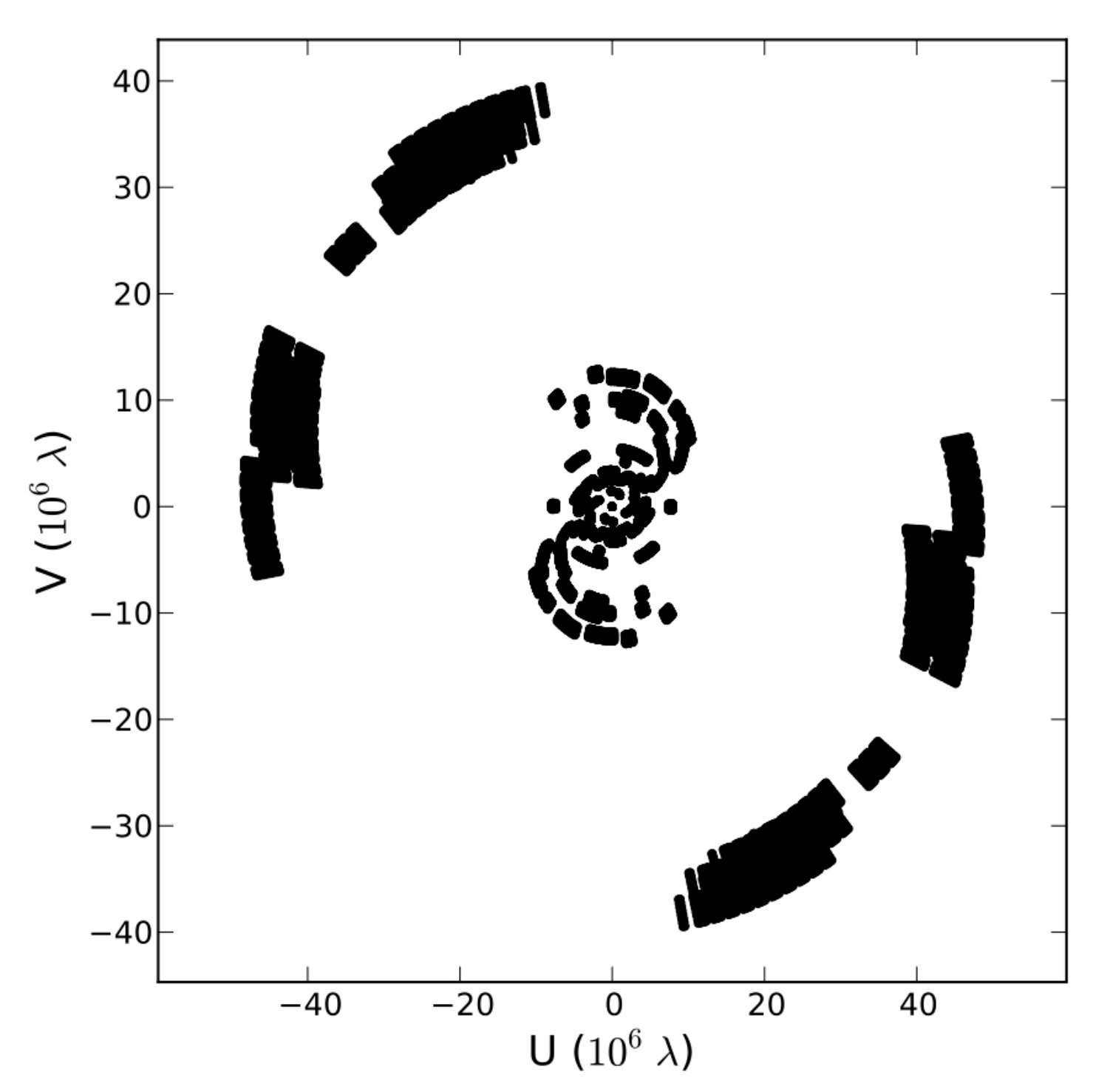}
\caption[uv-coverage of the SNR 4449-1]{{\em uv}-coverage for an integration time of 2 hours on SNR 4449-1.}
\label{uvcoverage}
\end{figure}

The remnant is clearly resolved into several ``spot-like'' components,
which is likely to reflect the limitations of deconvolution applied to
visibility data with an incomplete {\em uv}-coverage due to the short
integration time on the source (see Fig.~\ref{uvcoverage}). This
effect may also contribute to the apparent discrepancy between the
image in Fig.~\ref{fig1} and the structure observed with the
HSA\footnote{The High Sensitivity Array, comprised of 10 antennas of
  Very Long Baseline Array, the Robert C. Byrd Green Bank Telescope,
  and the Effelsberg 100 m telescope} at 1.4~GHz by 
\cite{2010MNRAS.409.1594B}. The total flux density and the spatial
extent of the radio emission are similar in both the 1.4~GHz HSA
image and the 1.6~GHz one.

The most likely factor contributing to the morphological discrepancy is the
difference of the baseline sensitivity and filling factors of the {\em
  uv}-coverages in the two data sets. The 1.4~GHz HSA data have a
better rms noise resulting from a longer integration time and a
superior {\em uv}-coverage on long baselines. However, the 1.6~GHz EVN
observations have better coverage and better sensitivity on baselines
shorter than 1000~km, which enhances detection
of emission on angular scales of $\sim$35 mas. Making a firm conclusion on the structure of this SNR would require obtaining a robust {\em uv}-coverage on baselines of $\le 1000$~km which can be provided by eMERLIN.

Six distinct regions (labelled A
through F in Fig.~\ref{fig1}) can be identified in the image. The
basic properties of components A--F (coordinates, total flux densities,
and brightness temperatures) have been obtained from fits by
two-dimensional Gaussian components to the image
plane. Table~\ref{table2} lists the parameters obtained. The errors
are taken from the covariance matrix of the fits.

The integrated flux density of the components identified varies from
0.269 to 0.648 mJy while the brightness temperature ranges from 
$3.0\times10^6$ to $5.6\times10^6$\,K. The whole structure has a
total flux of $6.1 \pm 0.6$\,mJy, from which we derive a radio
luminosity at 1.6 GHz of $1.74\times10^{35}$\,erg\,s$^{-1}$.

\begin{figure*}
\centering
\includegraphics[width=0.445\textwidth]{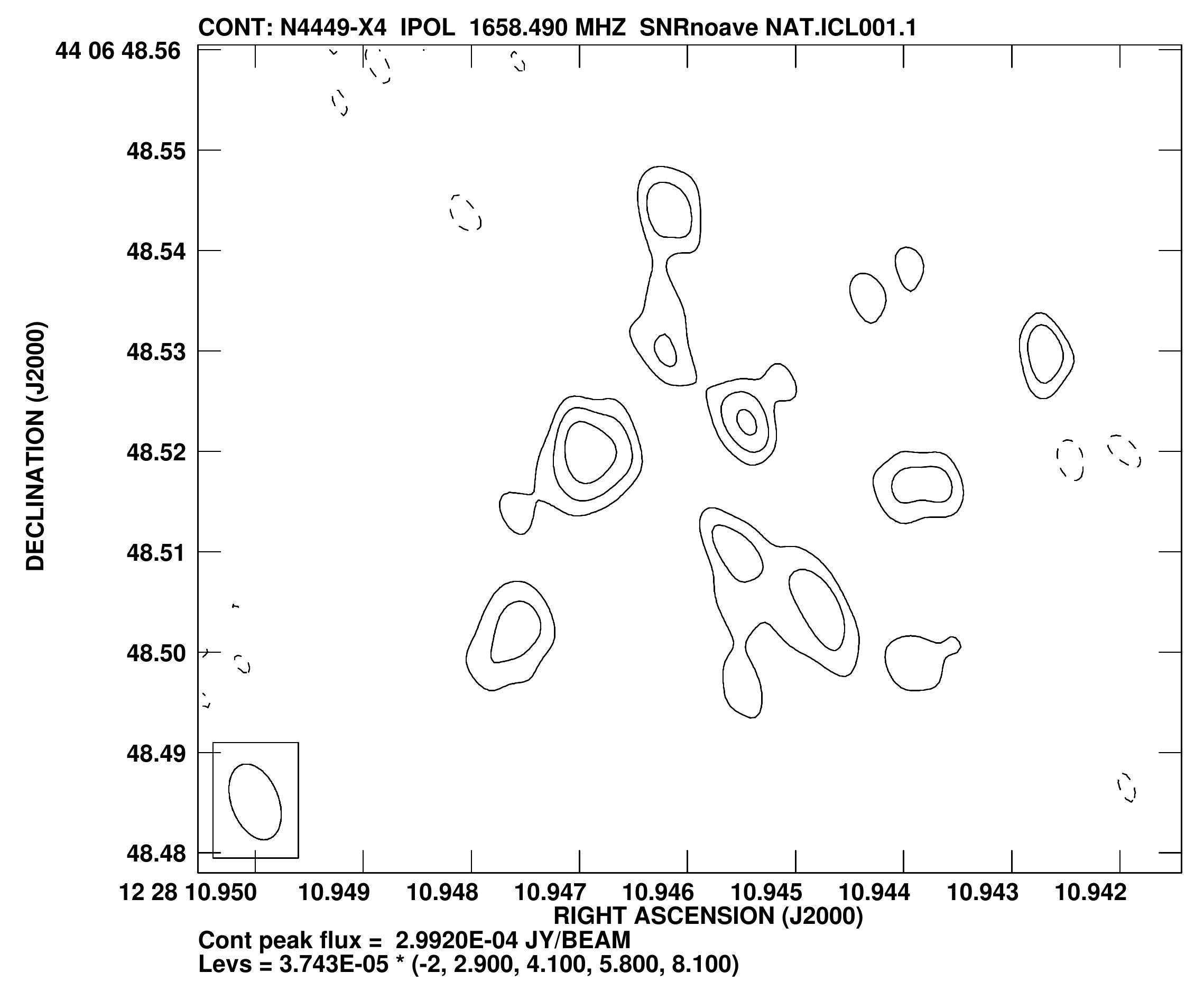}
\includegraphics[width=0.51\textwidth]{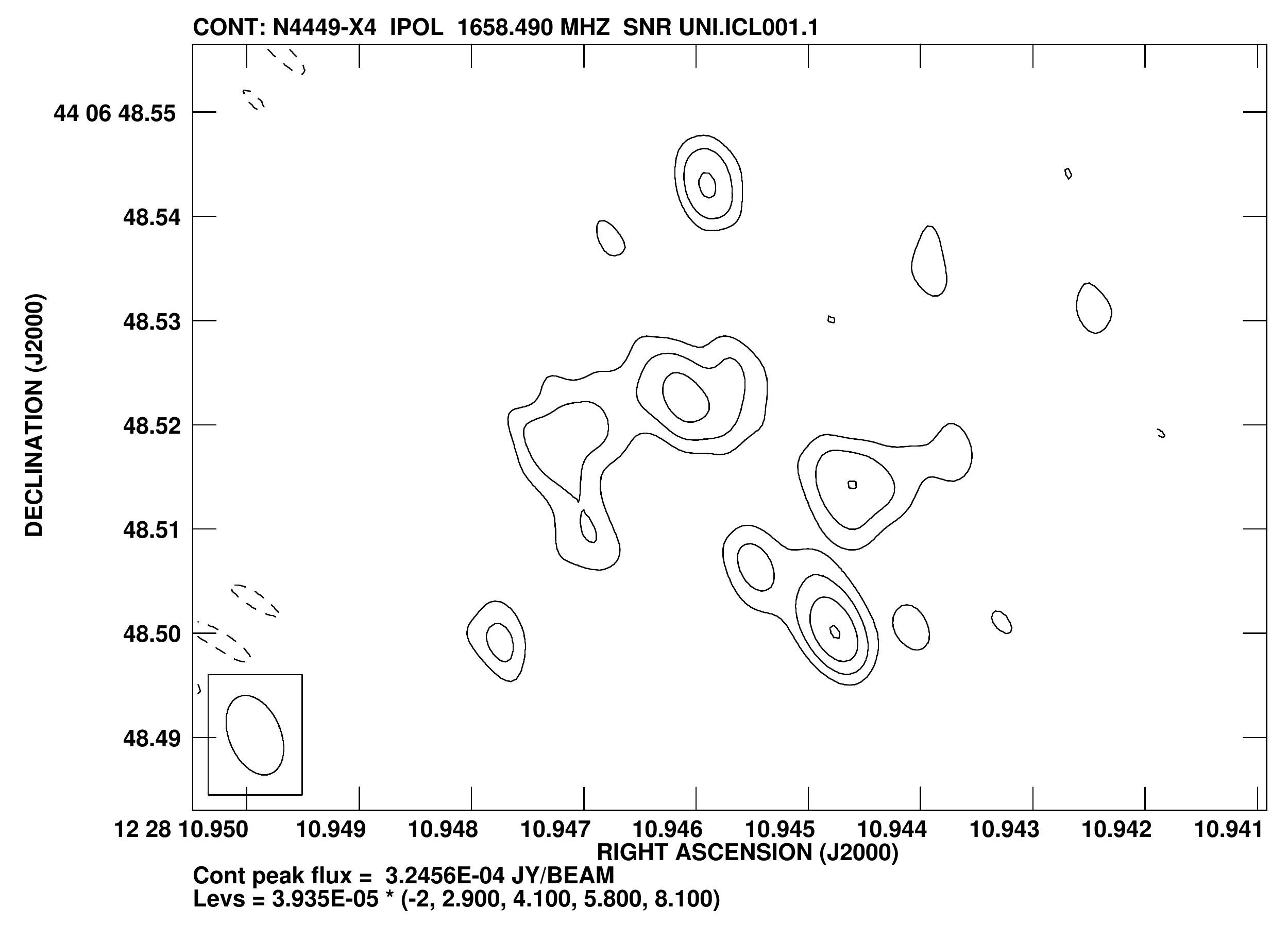}
\protect\caption{Resolved radio structure of the SNR in NGC\,4449 using non-frequency averaged data. Left: deconvolved image using natural weighting. The rms noise off-source is $37\mu$\,Jy\,beam$^{-1}$ and the size of the restoring beam is $7.9$\,mas$\times4.7$\,mas, at a position angle of 20$^{\circ}$.8. Right: deconvolved image using uniform weighting. The rms noise off-source is $39\mu$\,Jy\,beam$^{-1}$ and the size of the restoring beam is $8.0$\,mas$\times5.0$\,mas, at a position angle of 22$^{\circ}$.8. The contours start at $-1.5$ times the rms and increase with factors of $\sqrt{2}$. North is up and east is to the left.}
\label{fig1bis}
\end{figure*}

Component F is detected at an S/N$\sim5$, while an even fainter component of 0.150 mJy of integrated flux is also present in the image, to the north-west from component D. These components are likely to be spurious features resulting from a sidelobe or deficiencies of the CLEAN convolution. In order to check the reliability of such components, we investigate variations in the deconvolution and data weighting schemes. We first repeat the data calibration and, in order to improve the image fidelity, split the data without applying any frequency averaging. We then perform CLEAN deconvolution using natural weighting and uniform weighting. The resulting images have off-source rms noises of 37 and $39\,\mu$Jy beam$^{-1}$ (see Fig.~\ref{fig1bis}, left and right, respectively), and show a resolved structure formed by several spot-like components similar to those of Fig.~\ref{fig1}. Some of the components previously observed are not detected here (e.g., component A seems to have merged with B), while the previous possible spurious detection of component F seems to be still detected at an S/N$\sim5$. Despite the improvement in the image sensitivity, the structure of both images still shows an elliptical shape caused by the spatial frequencies sampled by the interferometer (see a further discussion on the image fidelity in the Appendix).

\begin{table*}
\protect\caption{Radio components of the SNR in NGC 4449}
\centering
\begin{tabular}{ccccc}
\hline
\hline 
 Component & RA & Dec. & Integrated flux & $T_\mathrm{B}$ \\
       & (J2000) & (J2000) & (mJy) & (10$^6$ K) \\
  \hline
  A & 12$^h$28$^m$10$^s$.94738 $\pm$ 0$^s$.00006 & 44$^{\circ}$06$^{\prime}$48$^{\prime\prime}$.5181 $\pm$ 0$^{\prime\prime}$.0008 & 0.524 & 3.8   \\
  B & 12$^h$28$^m$10$^s$.94691 $\pm$ 0$^s$.00006 & 44$^{\circ}$06$^{\prime}$48$^{\prime\prime}$.5200 $\pm$ 0$^{\prime\prime}$.0008 & 0.423 & 3.1   \\
  C & 12$^h$28$^m$10$^s$.94611 $\pm$ 0$^s$.00003 & 44$^{\circ}$06$^{\prime}$48$^{\prime\prime}$.5223 $\pm$ 0$^{\prime\prime}$.0004 & 0.648 & 5.6   \\
  D & 12$^h$28$^m$10$^s$.94444 $\pm$ 0$^s$.00005 & 44$^{\circ}$06$^{\prime}$48$^{\prime\prime}$.5146 $\pm$ 0$^{\prime\prime}$.0007 & 0.592 & 3.5   \\
  E & 12$^h$28$^m$10$^s$.94481 $\pm$ 0$^s$.00003 & 44$^{\circ}$06$^{\prime}$48$^{\prime\prime}$.5008 $\pm$ 0$^{\prime\prime}$.0005 & 0.424 & 4.5   \\
  F & 12$^h$28$^m$10$^s$.94588 $\pm$ 0$^s$.00004 & 44$^{\circ}$06$^{\prime}$48$^{\prime\prime}$.5423 $\pm$ 0$^{\prime\prime}$.0007 & 0.269 & 3.0   \\
\hline
\end{tabular}
\label{table2}
\end{table*}

\subsection{Positional centre and size of the remnant}
\begin{figure}
\centering
\includegraphics[width=\columnwidth]{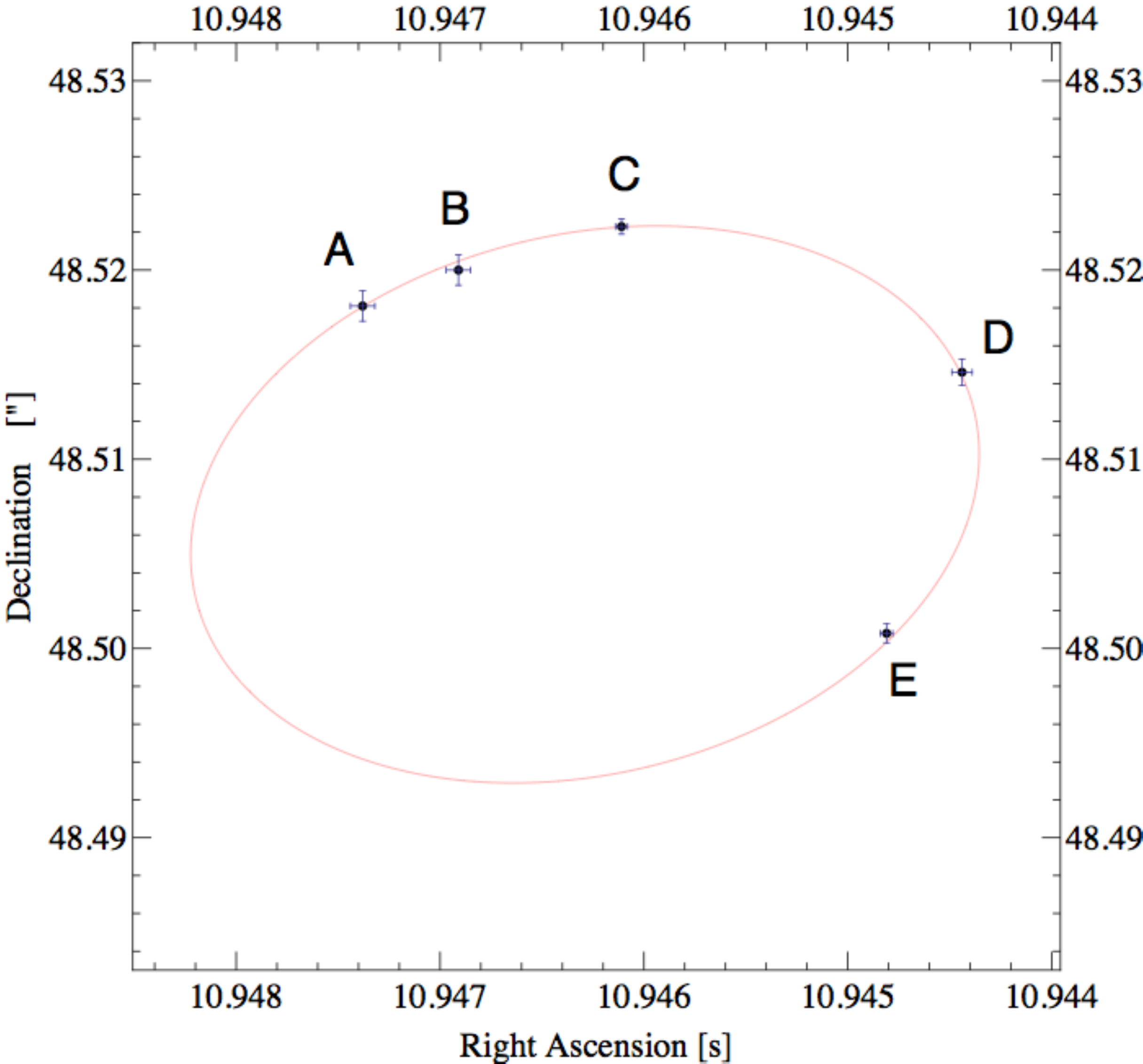}
\protect\caption{Elliptical fit to the peak of the five components detected. The ellipse has 
a major axis of 0.0422 arcsec $\pm$ 0.0023 arcsec, a minor axis of 
0.0285 arcsec $\pm$ 0.0024 arcsec, and a position angle of $119^{\circ}$. This corresponds to an ellipse of 0.8 pc $\times$ 0.5 pc.}
\label{fig2}
\end{figure}

The positional centre and an estimate of the SNR diameter size can be
determined from the resolved structure. The fit of an ellipse to the
structure formed by the peak of the five brightest components A-E (Fig.~\ref{fig2})
yields the best position for the SNR centre at\\
RA = 12$^h$28$^m$10$^s$.9463 $\pm$ 0$^s$.0001, Dec. =
44$^{\circ}$06$^{\prime}$48$^{\prime\prime}$.508 $\pm$ 0$^{\prime\prime}$.001,
and a SNR size of major axis $b$ = 0.0422 arcsec $\pm$
0.0023 arcsec and minor axis $a$ = 0.0285 arcsec $\pm$
0.0024 arcsec corresponding to 0.8 pc $\times$ 0.5 pc at the
distance of the galaxy. This estimate is limited by the uncertainty in
the distance to NGC 4449.  In order to compare the radio diameter to
the one derived from optical images, we derive a geometrical mean value of
0.0347 arcsec $\pm$ 0.0025 arcsec. This size is
larger than the value of 0.028 arcsec (\citealt{1998AAS...193.7404B}) and very similar to the 0.037 arcsec (\citealt{2008ApJ...677..306M})
estimated from the \textit{HST} images, suggesting that the
outer layers of SNR 4449--1 do also emit at optical wavelengths. The SNR
mean diameter of 35 mas is larger than the $\sim$30 mas
peak-to-peak separation between the two bright parallel ridges of
emission found by \cite{2010MNRAS.409.1594B}.    
If the component F also belongs to the remnant (as suggested in \citealt{2010MNRAS.409.1594B}), the size of the remnant can be determined from the distance between components A--D and F--E. This yields a size of 32 mas $\times$ 43 mas, agreeing better with the size obtained by \cite{2010MNRAS.409.1594B}.
It should be stressed here once more that deviations
from the expected circular or elliptical morphology reported by
\cite{2010MNRAS.409.1594B} for SNR 4449--1 are not likely
to result from deconvolution errors or gaps in the {\em uv}-coverage. These deviations (and the morphological discrepancy of the HSA and the EVN images of the remnant) most likely result from the
deficiency in the structural sensitivity of both data sets for scales larger than $\sim20$~mas. While peculiar non-circular structures have also been observed in other SNRs (e.g., 41.95+575 in M82; \citealt{2001MNRAS.322..100M}), the apparent discrepancy of the structural appearance of the SNR in our image and the image of \cite{2010MNRAS.409.1594B} calls for further more detailed imaging of this SNR.

\section{Discussion}
\subsection{Light curve}
\label{lightcurve}
\begin{figure}
\centering
\includegraphics[width=\columnwidth]{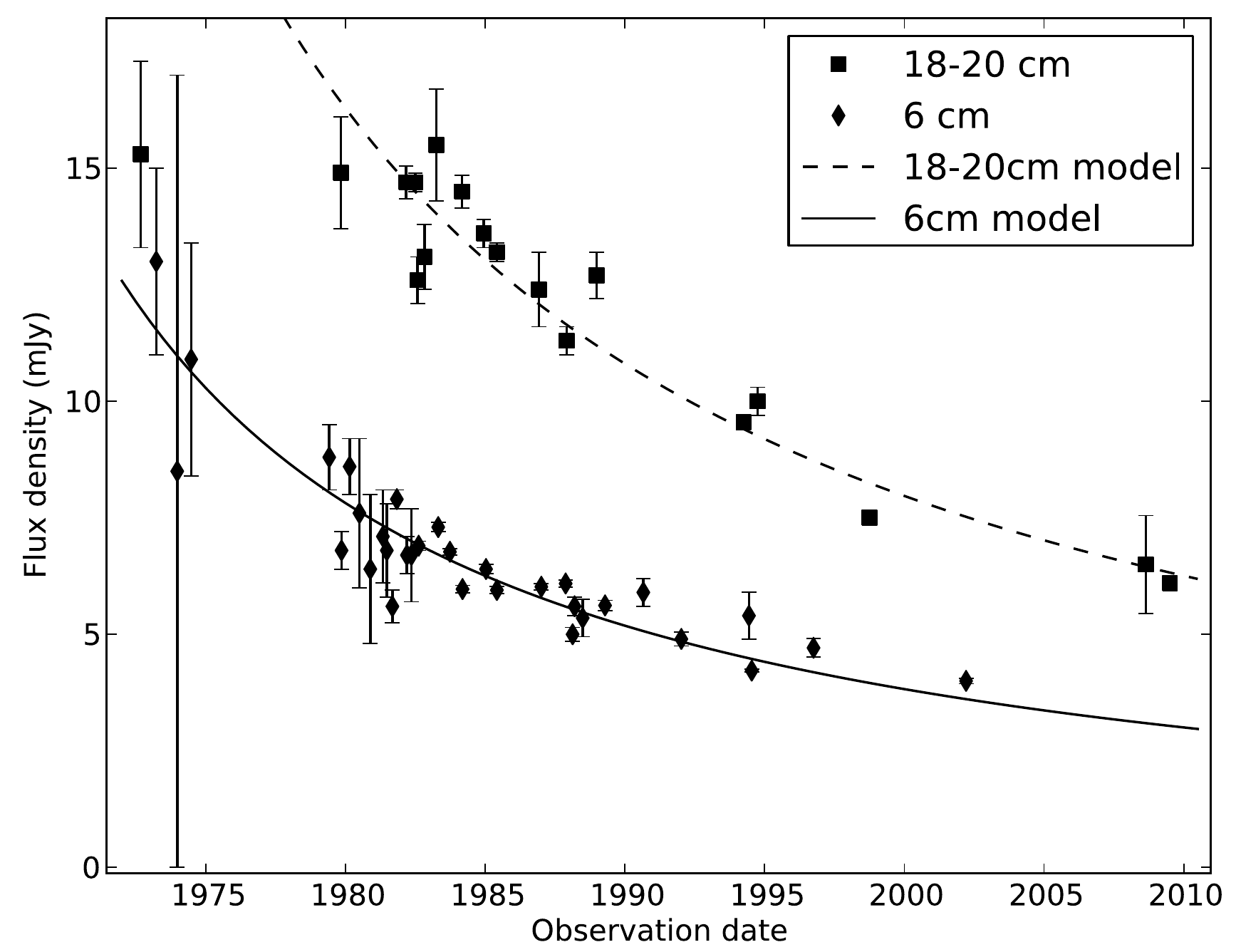}
\protect\caption{Lightcurve of SNR 4449--1 at 18 cm and 20 cm (squares), and 6 cm (diamonds). Data
are taken from Lacey et al. (2007), Bietenholz et al. (2010), 
and this work. The dashed line (thick line) represents our best-fitting model to the 18--20 cm (6 cm) 
data using equation~\ref{equation2}. Flux density measurements are treated separately at all three wavelengths. For presentation purposes, the 18 and 20 cm flux densities are plotted with the same symbol.}
\label{fig3}
\end{figure}

Using archival radio observations of NGC 4449, \cite{2007AJ....133.2156L} presented the light curves of the SNR at 6 and 20 cm
from 1972 to 2002 revealing a remarkable decline in its radio
emission. An up-to-date version of the light curves, including
the new flux densities from \cite{2010MNRAS.409.1594B} and this
work, is presented in Fig.~\ref{fig3}.

Our flux density measurement is in agreement with the radio
decline and we use it to estimate the decay rate of the emission from
a power-law fit of the form

\begin{equation} 
\label{equation2}
S = S_{0}(t-t_{0})^{\beta}\,\nu^{\alpha}\,,
\end{equation}
where $t_0$ is the date of the SN explosion, $S_{0}$ is a
scaling factor, $\nu$ is the observing frequency, $\alpha$ is the
spectral index, and $\beta$ is the index of the power-law decay (see,
e.g., \citealt{2002ARA&A..40..387W}). From a combined fit to the
lightcurves at 20, 18 and 6 cm (dashed and solid lines,
Fig.~\ref{fig3}), we obtain $S_\mathrm{0} = 922 \pm 307$\,Jy, $t_\mathrm{0} =
1956 \pm 2$ yr (AD), $\beta = -1.19 \pm 0.07$, and $\alpha = -0.620
\pm 0.004$, with a $\mathrm{\chi}^{2} = 9$.

The value of $\beta$ obtained from this fit yields a secular
decline rate of the radio emission of $2.2 \pm 0.1$\,\% per year.
This decline rate is similar to the ones found for SN 1970G in the
periods 1973--1991 and 1991--2001 (\citealt{1991ApJ...379L..49C}; \citealt{2001ApJ...559L.139S}), and it is slightly lower than the value
$2.8$\,\% obtained by \cite{2007AJ....133.2156L} under the assumption of
the SNR age of $\sim100$\,yrs.

\cite{2007AJ....133.2156L} report variations of the spectral
index, $\alpha$, observed over the time period 1972--2002. The spectral
index remains relatively constant from 1982 to 1996, with an average value of
$\alpha \approx$ $-$0.65 obtained largely from 1.4 and 4.8~GHz measurements.
This value is in good agreement with the spectral
index derived from our combined fit to the light curves at 6 and 20
cm. It steepens later to $\alpha \approx$ $-$1, albeit in measurements 
involving 22~GHz where spectral aging is more pronounced.

\subsection{Constraining the age of SNR\,4449--1}
\label{age}
The estimate of $t_{0}\sim1956$ yr (AD) indicates that the
SN should have exploded $\sim$55 yr ago.  This age agrees
with the upper limit of 100 years (\citealt{2008ApJ...677..306M}) and with the $\sim$70
yr suggested by \cite{2010MNRAS.409.1594B}. It is
also in agreement with the detections of the SNR on optical plates
(see \citealt{2008ApJ...677..306M} for a discussion),
which indicated that the SN explosion must have taken place in
1961 or before.

Using the [OIII] optical emission line of the SNR spectrum, a range of
expansion speeds of the SNR of 3000-6000 km\,s$^{-1}$ was found (\citealt{1998AAS...193.7404B}). An estimated maximum expansion velocity of
$\sim$6500 km s$^{-1}$ was later provided (\citealt{2010MNRAS.409.1594B}). This velocity range can be used to derive a
lower limit on the age of SNR 4449--1. Adopting the SNR radius of
14--21 mas, as given by the semimajor and semiminor axes of the
elliptical fit shown in Fig.~\ref{fig2} and an expansion speed of
$\sim$6500\,km\,s$^{-1}$, we estimate an age of the SNR between 39 and
59 yr (assuming a constant expansion speed). The SN
explosion would thus have taken place between 1951 and 1971, which is in
agreement with the $\mathrm{t_{0}}\sim1956$ found from the fit to the
radio light curves. Adopting an SNR radius of 16--21 mas, as given by 
the size estimate made with the feature F included, we obtain an age of the SNR of between 44 and 60 yr. This constrains the year of the SN explosion between 1950 and 1965, which is also in agreement with $\mathrm{t_{0}}\sim1956$.

We note that these estimates are made under the assumption that
the radio shell is expanding together with the optical [OIII] emitting
gas, at a constant speed of $\sim$6500 km s$^{-1}$. This is supported by the finding
that the size of the shell-like structure observed at 1.6 GHz is similar to the one in the optical. If the outer shock front were moving faster than 6500 km s$^{-1}$, the ages estimated using this expansion speed would constitute an upper limit to the remnant's age.

\subsubsection*{Decelerated expansion}
A deceleration in the shock front, produced by the interaction of the
ejecta with a dense CSM, would affect
the estimated age of the SNR  as well. \cite{1982ApJ...259L..85C,1982ApJ...259..302C} proposed that the relativistic
electrons and enhanced magnetic field necessary to produce the
observed radio synchrotron emission arise from the SN shock
wave interacting with a dense CSM, which is presumed to have been
established by a very effective mass-loss wind from the SN
stellar progenitor.

According to \cite{1982ApJ...259L..85C,1982ApJ...259..302C}, the shock front
of the SN in the CSM expands following a power law of time, $r
\propto t^{m}$, where $m$ is the {\em deceleration parameter} defined
as

\begin{equation}
\label{equation1}
m = (n - 3)/(n - s)\,.
\end{equation}

\noindent
The radial density profiles of the ejecta and the CSM are also
described as power laws, with indices $n$ and $s$, respectively (i.e.,
the ejecta density is $\propto r^{-n}$ and the CSM density is $\propto
r^{-s}$). The deceleration parameter, $m$, can also be written in
terms of the spectral index, $\alpha$, and the flux-density decay
rate, $\beta$, in the form (see \citealt{2011A&A...529A..47M}b)

\begin{equation}
\label{cooleq}
m = \frac{2(3+\alpha+\beta)}{3(s-4)+\alpha(s-2)}\,.
\end{equation}

\noindent
Taking the values of $\alpha = -0.62$ and $\beta = -1.19$, estimated
from our fit to the radio light curves, equation~\ref{cooleq} results in a
deceleration parameter $m = 0.80$, if we use a radial CSM
profile with $s = 2$ (i.e., if we assume a constant mass-loss rate for
the progenitor star). 
This value is in agreement with the typical values reported in
other SNe (where the expansion has been monitored during
several years). It is found in all cases that $0.7 \leq m \leq 0.9$
(\citealt{2002ApJ...581.1132B}; \citealt{2002ARA&A..40..387W}; \citealt{2009A&A...503..869M}; \citealt{2011A&A...526A.142M}a; and references therein).

\section{Summary}
We have presented the shell-like resolved structure of the young
oxygen-rich SNR in NGC 4449.  We obtain the most
accurate estimates of the SNR position and size (0.0422 arcsec $\times$ 0.0285 arcsec, corresponding to 0.8 pc $\times$ 0.5 pc at a distance of 3.82 Mpc),
making this object one of the largest extragalactic SNRs
imaged with VLBI. The historical light curve of the source can be well
fitted by a power law of index $\beta = -1.19 \pm 0.07$, after the
inclusion of the total flux density at 1.6 GHz, and an SNR age of
$\sim$55 yr. A decline rate of the radio emission of
$\mathrm{d}S/S\mathrm{d}t = -(2.2 \pm 0.1)$\,\% yr$^{-1}$ is found.

According to the fit of the radio light curves, the SN should
have exploded around 1956, which is in agreement with the detection of
the SN in the optical plates and with the age obtained assuming linear
expansion with a maximum speed derived from the optical oxygen lines
of $\sim$6500 km s$^{-1}$. The constraint of the explosion date of SNR 4449--1 places this source in a
unique position as being the youngest known supernova remnant (Cas A
is the youngest Galactic SNR with an age of $\sim$330 yrs; \citealt{2001AJ....122..297T}).

From the 1.6 GHz VLBI observation, we derive a luminosity of
$1.74\times10^{35}$ erg\,s$^{-1}$, which is comparable to the peak luminosity of
SN 1970G and higher than the current luminosities of the Galactic
SNRs Cas A and the Crab (see \citealt{2007AJ....133.2156L} for further
discussion). This makes SNR~4449--1 an interesting link between SNe and
their remnants and calls for classifying this object as a
transition-type source that links SN explosions and SNRs (\citealt{2007AJ....133.2156L}) or as an intermediate-age SNR with an exceptionally high radio luminosity (the so-called `radio hypernova remnant'; \citealt{1990MNRAS.242..529W}). 

Additional monitoring at multiple radio frequencies is required to
refine the fractional decline rate, light curve and spectral index
evolution of the SNR, and to determine its properties and
classification.

\section*{Acknowledgements}
The authors are grateful to the comments of the anonymous referee, which helped to improve the manuscript.
The EVN is a joint facility of European, Chinese,
South African and other radio astronomy institutes funded by their
national research councils. MM was supported for this research
through a stipend from the International Max Planck Research School
(IMPRS) for Astronomy and Astrophysics at the Universities of Bonn and
Cologne.

\bibliographystyle{mn2e} 
\bibliography{referencesALL}

\begin{thebibliography}{}

\bibitem[\protect\citeauthoryear{{Annibali}, {Aloisi}, {Mack}, {Tosi}, {van der
  Marel}, {Angeretti}, {Leitherer} \& {Sirianni}}{{Annibali}
  et~al.}{2008}]{2008AJ....135.1900A}
{Annibali} F.,  {Aloisi} A.,  {Mack} J.,  {Tosi} M.,  {van der Marel} R.~P.,
  {Angeretti} L.,  {Leitherer} C.,    {Sirianni} M.,  2008, \aj, 135, 1900

\bibitem[\protect\citeauthoryear{{Balick} \& {Heckman}}{{Balick} \&
  {Heckman}}{1978}]{1978ApJ...226L...7B}
{Balick} B.,  {Heckman} T.,  1978, \apjl, 226, L7

\bibitem[\protect\citeauthoryear{{Bietenholz}, {Bartel}, {Milisavljevic},
  {Fesen}, {Challis} \& {Kirshner}}{{Bietenholz}
  et~al.}{2010}]{2010MNRAS.409.1594B}
{Bietenholz} M.~F.,  {Bartel} N.,  {Milisavljevic} D.,  {Fesen} R.~A.,
  {Challis} P.,    {Kirshner} R.~P.,  2010, \mnras, 409, 1594

\bibitem[\protect\citeauthoryear{{Bietenholz}, {Bartel} \&
  {Rupen}}{{Bietenholz} et~al.}{2002}]{2002ApJ...581.1132B}
{Bietenholz} M.~F.,  {Bartel} N.,    {Rupen} M.~P.,  2002, \apj, 581, 1132

\bibitem[\protect\citeauthoryear{{Blair} \& {Fesen}}{{Blair} \&
  {Fesen}}{1998}]{1998AAS...193.7404B}
{Blair} W.~P.,  {Fesen} R.~A.,  1998, in BAAS. p.~1365

\bibitem[\protect\citeauthoryear{{Blair}, {Kirshner} \& {Winkler} Jr.}{{Blair}
  et~al.}{1983}]{1983ApJ...272...84B}
{Blair} W.~P.,  {Kirshner} R.~P.,    {Winkler} Jr. P.~F.,  1983, \apj, 272, 84

\bibitem[\protect\citeauthoryear{{Chevalier}}{{Chevalier}}{1982a}]{1982ApJ...259L..85C}
{Chevalier} R.~A.,  1982a, \apjl, 259, L85

\bibitem[\protect\citeauthoryear{{Chevalier}}{{Chevalier}}{1982b}]{1982ApJ...259..302C}
{Chevalier} R.~A.,  1982b, \apj, 259, 302

\bibitem[\protect\citeauthoryear{{Cowan}, {Goss} \& {Sramek}}{{Cowan}
  et~al.}{1991}]{1991ApJ...379L..49C}
{Cowan} J.~J.,  {Goss} W.~M.,    {Sramek} R.~A.,  1991, \apjl, 379, L49

\bibitem[\protect\citeauthoryear{{de Bruyn}}{{de
  Bruyn}}{1983}]{1983A&A...119..301D}
{de Bruyn} A.~G.,  1983, \aap, 119, 301

\bibitem[\protect\citeauthoryear{{Eck}, {Roberts}, {Cowan} \& {Branch}}{{Eck}
  et~al.}{1998}]{1998ApJ...508..664E}
{Eck} C.~R.,  {Roberts} D.~A.,  {Cowan} J.~J.,    {Branch} D.,  1998, \apj,
  508, 664

\bibitem[\protect\citeauthoryear{{Heywood}, {Blundell}, {Kl{\"o}ckner} \&
  {Beasley}}{{Heywood} et~al.}{2009}]{2009MNRAS.392..855H}
{Heywood} I.,  {Blundell} K.~M.,  {Kl{\"o}ckner} H.-R.,    {Beasley} A.~J.,
  2009, \mnras, 392, 855

\bibitem[\protect\citeauthoryear{{Lacey}, {Goss} \& {Mizouni}}{{Lacey}
  et~al.}{2007}]{2007AJ....133.2156L}
{Lacey} C.~K.,  {Goss} W.~M.,    {Mizouni} L.~K.,  2007, \aj, 133, 2156

\bibitem[\protect\citeauthoryear{{Liu} \& {Bregman}}{{Liu} \&
  {Bregman}}{2005}]{2005ApJS..157...59L}
{Liu} J.-F.,  {Bregman} J.~N.,  2005, \apjs, 157, 59

\bibitem[\protect\citeauthoryear{{Marcaide}, {Mart{\'{\i}}-Vidal},
  {Perez-Torres}, {Alberdi}, {Guirado}, {Ros} \& {Weiler}}{{Marcaide}
  et~al.}{2009}]{2009A&A...503..869M}
{Marcaide} J.~M.,  {Mart{\'{\i}}-Vidal} I.,  {Perez-Torres} M.~A.,  {Alberdi}
  A.,  {Guirado} J.~C.,  {Ros} E.,    {Weiler} K.~W.,  2009, \aap, 503, 869

\bibitem[\protect\citeauthoryear{{Mart{\'{\i}}-Vidal}, {Marcaide}, {Alberdi},
  {Guirado}, {P{\'e}rez-Torres} \& {Ros}}{{Mart{\'{\i}}-Vidal}
  et~al.}{2011}]{2011A&A...526A.142M}
{Mart{\'{\i}}-Vidal} I.,  {Marcaide} J.~M.,  {Alberdi} A.,  {Guirado} J.~C.,
  {P{\'e}rez-Torres} M.~A.,    {Ros} E.,  2011, \aap, 526, A142

\bibitem[\protect\citeauthoryear{{Mart{\'{\i}}-Vidal}, {P{\'e}rez-Torres} \&
  {Brunthaler}}{{Mart{\'{\i}}-Vidal} et~al.}{2011}]{2011A&A...529A..47M}
{Mart{\'{\i}}-Vidal} I.,  {P{\'e}rez-Torres} M.~A.,    {Brunthaler} A.,  2011,
  \aap, 529, A47

\bibitem[\protect\citeauthoryear{{McDonald}, {Muxlow}, {Pedlar}, {Garrett},
  {Wills}, {Garrington}, {Diamond} \& {Wilkinson}}{{McDonald}
  et~al.}{2001}]{2001MNRAS.322..100M}
{McDonald} A.~R.,  {Muxlow} T.~W.~B.,  {Pedlar} A.,  {Garrett} M.~A.,  {Wills}
  K.~A.,  {Garrington} S.~T.,  {Diamond} P.~J.,    {Wilkinson} P.~N.,  2001,
  \mnras, 322, 100

\bibitem[\protect\citeauthoryear{{Mezcua}, {Farrell}, {Gladstone} \&
  {Lobanov}}{{Mezcua} et~al.}{2013}]{2013arXiv1309.4463M}
{Mezcua} M.,  {Farrell} S.~A.,  {Gladstone} J.~C.,    {Lobanov} A.~P.,  2013,
  MNRAS, preprint (arXiv:1309.4463)

\bibitem[\protect\citeauthoryear{{Mezcua} \& {Lobanov}}{{Mezcua} \&
  {Lobanov}}{2011}]{2011AN....332..379M}
{Mezcua} M.,  {Lobanov} A.~P.,  2011, Astron. Nachr., 332, 379

\bibitem[\protect\citeauthoryear{{Milisavljevic} \& {Fesen}}{{Milisavljevic} \&
  {Fesen}}{2008}]{2008ApJ...677..306M}
{Milisavljevic} D.,  {Fesen} R.~A.,  2008, \apj, 677, 306

\bibitem[\protect\citeauthoryear{{Pennington} \& {Dufour}}{{Pennington} \&
  {Dufour}}{1983}]{1983ApJ...270L...7P}
{Pennington} R.~L.,  {Dufour} R.~J.,  1983, \apjl, 270, L7

\bibitem[\protect\citeauthoryear{{Reines}, {Johnson} \& {Goss}}{{Reines}
  et~al.}{2008}]{2008AJ....135.2222R}
{Reines} A.~E.,  {Johnson} K.~E.,    {Goss} W.~M.,  2008, \aj, 135, 2222

\bibitem[\protect\citeauthoryear{{S{\'a}nchez-Sutil}, {Mu{\~n}oz-Arjonilla},
  {Mart{\'{\i}}}, {Garrido}, {P{\'e}rez-Ram{\'{\i}}rez} \&
  {Luque-Escamilla}}{{S{\'a}nchez-Sutil} et~al.}{2006}]{2006A&A...452..739S}
{S{\'a}nchez-Sutil} J.~R.,  {Mu{\~n}oz-Arjonilla} A.~J.,  {Mart{\'{\i}}} J.,
  {Garrido} J.~L.,  {P{\'e}rez-Ram{\'{\i}}rez} D.,    {Luque-Escamilla} P.,
  2006, \aap, 452, 739

\bibitem[\protect\citeauthoryear{{Seaquist} \& {Bignell}}{{Seaquist} \&
  {Bignell}}{1978}]{1978ApJ...226L...5S}
{Seaquist} E.~R.,  {Bignell} R.~C.,  1978, \apjl, 226, L5

\bibitem[\protect\citeauthoryear{{Stockdale}, {Goss}, {Cowan} \&
  {Sramek}}{{Stockdale} et~al.}{2001}]{2001ApJ...559L.139S}
{Stockdale} C.~J.,  {Goss} W.~M.,  {Cowan} J.~J.,    {Sramek} R.~A.,  2001,
  \apjl, 559, L139

\bibitem[\protect\citeauthoryear{{Summers}, {Stevens}, {Strickland} \&
  {Heckman}}{{Summers} et~al.}{2003}]{2003MNRAS.342..690S}
{Summers} L.~K.,  {Stevens} I.~R.,  {Strickland} D.~K.,    {Heckman} T.~M.,
  2003, \mnras, 342, 690

\bibitem[\protect\citeauthoryear{{Thorstensen}, {Fesen} \& {van den
  Bergh}}{{Thorstensen} et~al.}{2001}]{2001AJ....122..297T}
{Thorstensen} J.~R.,  {Fesen} R.~A.,    {van den Bergh} S.,  2001, \aj, 122,
  297

\bibitem[\protect\citeauthoryear{{Weiler}, {Panagia}, {Montes} \&
  {Sramek}}{{Weiler} et~al.}{2002}]{2002ARA&A..40..387W}
{Weiler} K.~W.,  {Panagia} N.,  {Montes} M.~J.,    {Sramek} R.~A.,  2002,
  \araa, 40, 387

\bibitem[\protect\citeauthoryear{{Wilkinson} \& {de Bruyn}}{{Wilkinson} \& {de
  Bruyn}}{1990}]{1990MNRAS.242..529W}
{Wilkinson} P.~N.,  {de Bruyn} A.~G.,  1990, \mnras, 242, 529

\bibitem[\protect\citeauthoryear{{Yokogawa}, {Imanishi}, {Koyama}, {Nishiuchi}
  \& {Mizuno}}{{Yokogawa} et~al.}{2002}]{2002PASJ...54...53Y}
{Yokogawa} J.,  {Imanishi} K.,  {Koyama} K.,  {Nishiuchi} M.,    {Mizuno} N.,
  2002, \pasj, 54, 53

\end{thebibliography}

\section*{Appendix}
\subsection*{Discussion on the image fidelity}
The sparse \textit{uv}-coverage of our observations has a direct effect on the fidelity of the SNR N4449$-$1 image. Holes in the \textit{uv}-coverage map into a complete filtering of a range of spatial frequencies in the image plane. Hence, any structure with spatial scales corresponding to spatial frequencies not covered in our observations will be invisible to the interferometer.

On the one hand, extended sources (like spherical shells or rings) have different amplitude components at different spatial scales. If we are unable to measure all these components at all spatial frequencies, it is difficult (if not impossible) to recover the source structure with a high fidelity, unless some assumptions are made. 
On the other hand, things are different for sources consisting of a discrete set of compact (i.e., point-like) components. A point source has exactly the same amplitude at all the spatial scales. Hence, the imaging of such a source (or a discrete set of such sources) is very robust, even if the \textit{uv}-coverage is sparse. 

Our image of SNR N4449$-$1 consists of a set of compact sources distributed following a somewhat elliptical manner. This image can be understood as an elliptical structure, whose spatial frequencies have been filtered by a sparse \textit{uv}-coverage. As a result of the spatial-frequency filtering, the image is only sensitive to the portions of the elliptical structure whose spatial frequencies are sampled by the interferometer. The effect of a sparse \textit{uv}-coverage on extended sources (in particular, the shell-like structures corresponding to radio-SNRs) was studied by \cite{2009MNRAS.392..855H} and discussed (and discarded) in \cite{2010MNRAS.409.1594B} as a possible origin of the distorted structure seen in their HSA image of SRN N4449$-$1. 

\begin{figure}
\centering
\includegraphics[width=\columnwidth]{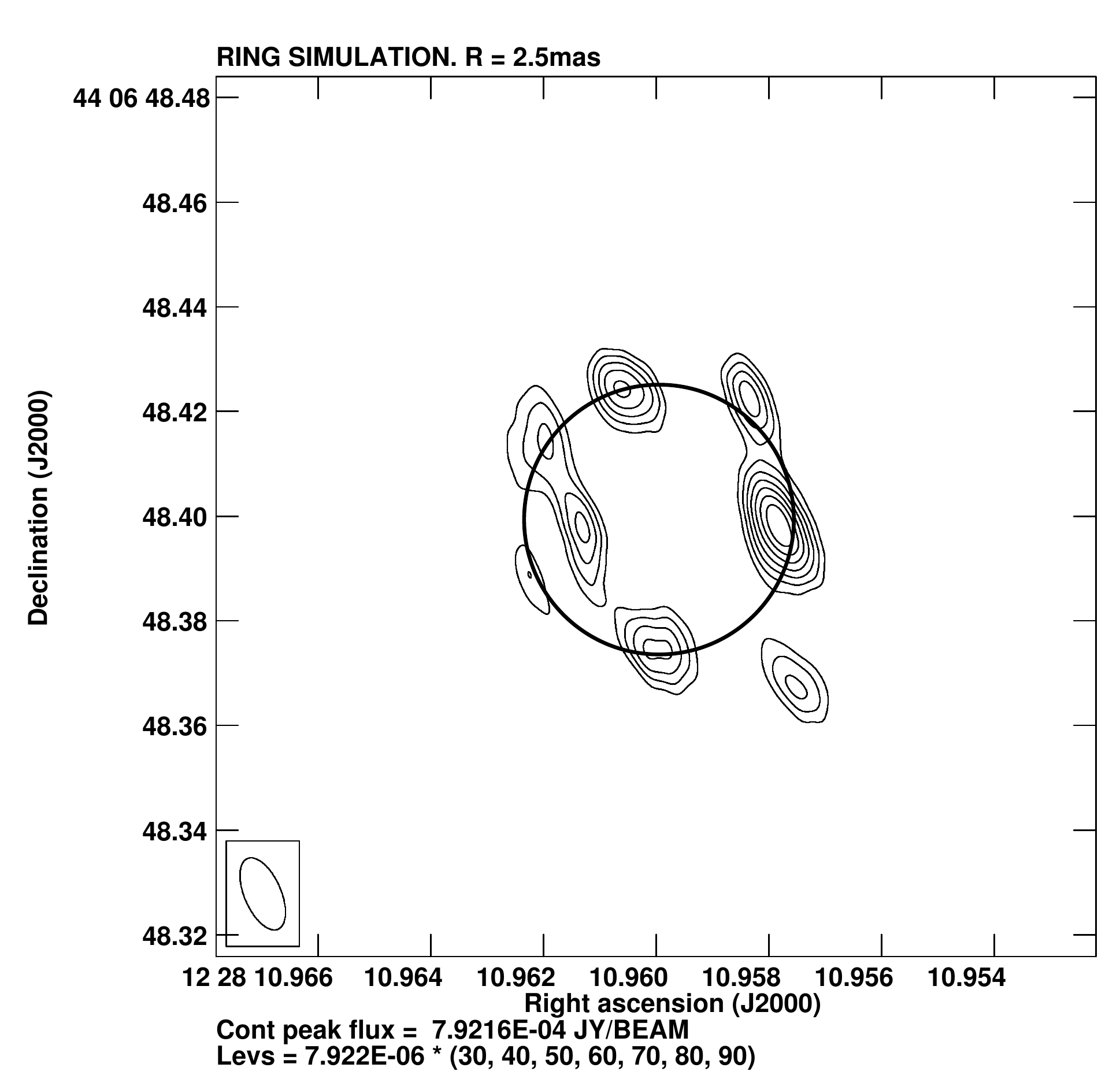}
\protect\caption{Contour plot of the deconvolved image of a spherical shell, of size 50\,mas, obtained from a realistic simulation of our EVN observations (see the text). The thick circle line marks the outer radius of the simulated shell.}
\label{SimulShell}
\end{figure}

We have performed a test similar to that described in \cite{2010MNRAS.409.1594B}, although with some substantial differences. Instead of simulating observations of a spherical shell from a synthetic \textit{uv}-coverage and receiver noise, we added the shell model to the residual (i.e., post-CLEAN) visibilities of our own observations. This way, the noise and the visibility weights result to be {\em exactly} the same as those of the real source observations. The deconvolved image of the spherical shell (of size 50\,mas) is shown in Fig. \ref{SimulShell}.The recovered signal from the shell is broken into several compact features, which are distributed around the outer edge of the source structure. Hence, if the structure of SNR N4449$-$1 is extended (as it is indeed the case), our sparse \textit{uv}-coverage translates into an incomplete detection of the whole structure, with several compact components distributed along the radio-emitting region. Hence, it is not possible to discern the innermost details of the source structure from our observations (i.e., we cannot tell whether the source resembles more a disc, a ring, or a shell), but we can still get information on the source size, based on the distribution of the compact components recovered in our image.

\label{lastpage}

\end{document}